# *knowCC*: Knowledge, awareness of computer & cyber ethics between CS/non-CS university students

Naresh Kshetri
Dept. of Accounting & Technology
Emporia State University
Emporia, Kansas, USA, 66801
NKshetri@emporia.edu

Vasudha
Department of CSE
United Institute of Management
Prayagraj, UP, INDIA, 211010
vas.is.2008@gmail.com

Denisa Hoxha
Department of Math, CS, & IT
Lindenwood University
St. Charles, Missouri, USA, 63301
DH770@lindenwood.edu

*Abstract*—Technology has advanced dramatically in the previous several years. There are also cyber assaults. Cyber-attacks pose a possible danger to information security and the general public. Since data practice and internet consumption rates continue to upswing, cyber awareness has become progressively important. Furthermore, as businesses pace their digital transformation with mobile devices, cloud services, communal media, and Internet of Things services, cybersecurity has appeared as a critical issue in corporate risk management. This research focuses on the relations between cybersecurity awareness, cyber knowledge, computer ethics, cyber ethics, and cyber behavior, as well as protective tools, across university students in general. The findings express that while internet users are alert of cyber threats, they only take the most elementary and easy-to-implement precautions. Several knowledge and awareness have been proposed to knob the issue of cyber security. It also grants the principles of cybersecurity in terms of its structure, workforces, and evidence pertaining to the shield of personal information in the cyber world. The first step is for people to educate themselves about the negative aspects of the internet and to learn more about cyber threats so that they can notice when an attack is taking place. To validate the efficiency of the suggested analysis between CS and non-CS university students, case study along with several comparisons are provided.

## I. INTRODUCTION

Cyber defense and cybersecurity demonstrate that higher levels of cyber knowledge are related to higher levels of cyber awareness, irrespective of respondent homeland or gender. Moreover, awareness is linked to security tools, but not to the information they are prepared to provide, personal characters to predict cyber behaviors [1]. Businesses are striving to engage their consumers by building social commerce websites, which give clients a unique buying experience, due to the tremendous development in the number of individuals utilizing social networking sites, cyber awareness in online education process is must [2]. Hence, even if someone is uncomfortable with having their private information exposed on the internet, they have no other option. Because everyone has some information online at some time, cyber attackers are attracted to begin illicit operations that are deemed theft.

Cybersecurity is a universal phenomenon that poses a difficult socio-technical challenge to regimes while also needing the participation of citizens [3]. Although cybersecurity is one of the most pressing issues threatening governments today, civic knowledge and visibility remain low.

Although almost all has heard about cybersecurity, people's earnestness and behavior do not mirror a high level of understanding, growth of online devices, users, shopping, gaming, teaching has given upsurge to online crimes [4]. Because the population has a skeptical view and understanding of these major threats, it makes it easier for attackers to take action in secret without the victim knowing. Apart from the risk of losing your information and the attacker profiting financially, there are further concerns. In certain circumstances, people's private social media accounts are stolen, and the hacker uses their personal information to blackmail them in return for money, and the victim is stuck for as long as the attacker wishes. Another major issue is that innocent children are also being victims. Because the population has a skeptical view and understanding of these major threats, it makes it easier for attackers to take action in secret without the victim knowing. There are paradoxes in communicating cybersecurity, which has led to society failing to take sufficient actions to cope with the risks.

The article is organized into different sections as related work in Section 2 which discusses mainly about the various articles which highlight the knowledge and awareness of cyber security at different domains of the society. We then presented Section 3 of our paper which contains the Case Study performed on the university students based on some common questions for analyzing their awareness and knowledge towards cyber security. In Section 4 we performed the Analysis on the data collected from Section 3, in this we compared the data with respect to one another for gaining the percentage of knowledge towards the cybersecurity. Then our Section 5 highlights the need of Cyber awareness which can help in reducing the level of threats occurring while using the internet. Section 6 concluded all the important points which are required for increasing the cyber awareness among the University students. In Section 7, we gave the future scope of the study as well as offered the key areas of improvement in the future research.

.

## II. LITERATURE REVIEW

In [1], SM Kennison et al. (2020), investigated how well self-reported unsafe cybersecurity behavior could be forecasted by a combination of self-reported knowledge about secure passwords and personal characteristics, such as personality attributes and general risk-taking in everyday life. The study is conducted on 325 undergraduate's participants and their self-reported unsafe cybersecurity behaviors which include using not secured Wi-Fi connections, not properly logging out their accounts on public computers, use of weak passwords and five big personality traits. The main focus of the author to highlight the superior levels of mood instability predicted superior levels of self-reported unsafe cyber security behavior and any one can be the victim with respect to cybersecurity, it is dissimilar to individual differences in personality.

In [2], ME Erendor et al. (2022), case study analysis is conducted to analyze the knowledge level of Kyrgyz - Turkish Manas University scholars about cybersecurity awareness in the distance education process. For this a survey is conducted on 517 students which include graduates, postgraduates and PhD scholars are asked certain questions focusing on malicious software, password safety and social media safety. The result of the report shows that after much use of digital platforms as well as internet technology the students still lack cyber security knowledge and have to be guided for the use of the internet more efficiently.

In [3], Herath, T. B. G., Khanna, P., & Ahmed, M. (2022), article points out the cybersecurity on Social Media which is a set of electronic interaction channels that internet users utilize to construct online communities. The authors aimed to determine the factors influencing users' knowledge of security-related features on social media platforms, as well as the influence of social media users' alertness on their behavior on social media platforms. Many cyber threats have been discovered within social media platforms, including cost of productivity, hijacking, identity theft, inconsistent personal branding, individual reputational damage, sexual harassments, data breach, malicious software, hacks, cyberbullying, password cracking, fake accounts, malware attacks, Sybil attacks, spear phishing attacks, and so on. The majority of survey participants used Facebook in their everyday lives to create associations with friends and family, play games, read articles, and access audio and video clips.

In [5], Alharbi, T. & Tassaddiq A. (2021), Most students participate in data breaches and numerical misbehavior owing to a lack of cybersecurity education and awareness, as well as the repercussions of cybercrime. The purpose of this study was to explore and assess the level of cybersecurity knowledge and user obedience among Majmaah University undergraduate students using a precise questionnaire based on many internet security aspects. To get the study's results, a quantitative research approach was employed, as well as several statistical tests, such as ANOVA, Kaiser-Meyer-Olkin (KMO), and Bartlett's tests, to assess and investigate the hypotheses. This research thoroughly evaluated safety problems for electronic emails, CPU viruses, phishing, falsified adverts, popup windows, and other Internet outbreaks. Ultimately, recommendations based on the obtained data are offered to cope with this widespread problem. According to the authors, educational institutions should offer comprehensive security awareness and training sessions on a routine basis to guarantee that all users are alert of the most repeated cybersecurity risks and vulnerabilities.

In [6], Senthilkumar, K. & Easwaramoorthy S. (2017), survey aims to investigate cyber security awareness among college students in Tamil Nadu by concentrating on various safety dangers on the internet. It is claimed that females are the most common victims of cybercrime. An essential part of the paper is the solution paragraph. There are essential elements that emphasize uses to frame the awareness on User ID and password, identifying knowledge on Home Computer Protection, firewall installation information, and the requirement for Antivirus software, and so on. According to study results, more than 70% of students from all cities are aware of basic viral stabbings and use antivirus software (which is constantly updated) or Linux platforms to shield their systems from virus attack. Every student is a member of one or more social networks. As a result, the survey is being done to determine the quantity of personal data that each student publishes on social media too.

In [7], N. Kshetri et al. (2022), surveyed blockchain technology applications for cyber defense and cybersecurity by summarizing blockchain architecture and models used. Cyber defense is an organized act of resistance or defense mechanism developed for information, system, and network security in contradiction of offensive cyber operations by implementing security properties. The authors also presented countermeasures with the help of blockchain technology and analyzed benefits of blockchain technology w.r.t. cyber defense and cyber security. Before comparing the solutions domains including challenges of blockchain and type of blockchain used, authors have pointed out and discussed the underlying properties of blockchain (decentralization, immutability, traceability, privacy, consensus mechanism) that provide a robust base for cyber defense and security of data in cyberspace.

In [8], Zwilling, M. et. al. (2020). Article evaluates the threats that are represented to cyber security. The study links cyber hazard awareness and risk exposure to certain user characteristics (gender, age, degree of IT use, etc.), concluding that customized training programs should be designed by educational and academic institutions. While this is a preliminary study, the authors want to utilize a comparative technique to assess cultural variations in cyber security awareness, knowledge, and practices. Forthcoming research should, however, focus on the causes of this lack of cyber threat awareness. According to the article, there is an economic necessity for industrialized nations with high GDP values (such as Israel) to invest in cyber security technologies because much of the population lacks the required skills and knowledge to guard against cyber risks. Yet, it is critical to invest in cyber training in order to influence people's perceptions about cyber threats. The internet is fully established in our daily lives, and our dependance on linked mobile devices seems to be increasing, which appears to be a major problem.

In [9], Chandarman, R. & Van Niekerk, B. (2017)) this article argues the growth in the number of cases of cyber-attacks among the population is due to the absence of cybersecurity awareness among the peoples. The lack of awareness may enhance the chance of cyber-attacks, so the author pointed out the need for cybersecurity awareness (CSA) and it must be incorporated into organizational policies and standards. For the study the author takes the data from the students on specific domains like key security, cyberbullying, phishing, malware, identity robbery and downloading, contribution and use of pirated substance and this collected data is then analyzed on the specific parameters like how much they have knowledge about the threats, self-perception of cybersecurity talents, actual cybersecurity skills and performance and correct cybersecurity manner. The last author finds that the number of cybercrime victims can be abridged by increasing the cybersecurity awareness (CAS) among the peoples and this awareness must be implemented in the detailed spectrum.

In [10], Kovacevic, A. et. al. (2020) conducted survey among university scholars in Serbia (in particularly freshman), which evaluates the factors relevant for cyber security awareness in strength. The survey also scrutinized unreported correlations; how various factor in particular and jointly, such as socio-demographic features, cyber security acuities, cyber security breach familiarities, IT usage, and knowledge impact security behavior. Nobody of the participants resolved all of the questions correctly in the part of the questionnaire concerning knowledge, which escorted to the assumption that students do not entertain the required knowledge or acceptable awareness of threats in cyber world.

In [11], Chaplinska, I. (2019). There are numerous challenges that past generations did not encounter in today's society, which is plagued with technological breakthroughs and devices, virtual communities, and dating, para-social contact, and robotics. In the essay, incidents are presented in which adolescent girls, uninformed of the risks, trust their intimate images with untrustworthy people or kidnap personal information before blackmailing or harassing them. There have also been incidents where minors have trusted social networks with details about their location, family composition, and property wealth, only to face fraud or physical harm. Such incidents have demonstrated that the culture of cyber security has not been instilled in the Ukrainian people's young generation, which has become a major issue. For example, at the Kyiv School, this has already occurred. A group of teens harassed one of their classmates by posting pornographic picture collages on social media. Children used photoshop to add the heads of their classmates to images of pornographic actresses.

III. CASE STUDY OF UNIVERSITY STUDENTS

Cybersecurity knowledge is more important than ever in the modern digital world. It is crucial to be aware of the possible hazards and threats linked with the use of technology and the internet since cyber assaults are becoming more frequent and sophisticated. Understanding the various attack types, such as phishing, malware, and ransomware, and being able to spot and prevent them are all part of cybersecurity awareness. Best practices for password management, data security, and network security must also be understood. Strong cybersecurity knowledge is crucial for both individuals and companies because it helps secure sensitive information, stop cyberattacks, and lessen the effects of any potential security catastrophes for business [12]. Individuals may contribute to ensuring a safe and secure digital environment for themselves and others by remaining educated and taking a proactive attitude to cybersecurity.

Q1: How frequently does the respondent back up crucial data and files? Answers include: every day, every week, every month, and never. Q2 asks if the respondent utilizes two-factor authentication for crucial accounts like banking or email. There are just two possible responses: yes or no. Q3 inquires as to whether the respondent has ever obtained a software or file from an unreliable source. There are three possible responses: yes, no, or maybe. Q4 asks if the respondent upgrades the operating system and software on their devices on a frequent basis. There are just two possible responses: yes or no. Q5 inquires as to how many hours a day the respondent spends online. 1 to 5 hours, 6 to 10 hours, or 10+ hours are the available alternatives for replying. Q6:The answer to this question will indicate if the respondent thinks that both people and organizations, or both, are responsible for cybersecurity. You have the choice of responding on behalf of people, organizations, or both. Q7: Whether through online or offline means, how does the person pay their power bills? Q 8: Did the person take any online courses as part of their degree program? There are just two possible responses: yes or no. Q9. This question asks if the respondent utilizes Wi-Fi in public spaces like restaurants, malls, or airports. There are just two possible responses: Yes or No. Q10 inquires as to how frequently the respondent changes passwords. There are four possible responses: often, monthly, yearly, and weekly.

In order to gather data, a survey about cybersecurity awareness and ethics was given to 150 students at University level. The survey's questions are intended to learn more about the students' understanding of, attitudes about, and conduct around cybersecurity awareness and ethics. A sample of students is electronically given the survey as part of the data collection procedure. Depending on the goals and objectives of the research, a random selection methodology or a stratified sample method is used to choose the students. The students were chosen based on their majors: *computer science*, *IT or cybersecurity*, and *non-computer or non-IT or non-cybersecurity* majors. To make inferences about the students' cybersecurity ethics after analyzing the survey replies, the data first is gathered. The data is then statistically analyzed, such as by computing percentages or frequencies. A report or presentation that includes tables, charts, and other visual aids is created to assist convey the results. A suggestion or an inference based on the study's findings is also included in the study with more social and ethical challenges sooner or later with both computer and cyber [13].

IV. ANALYSIS OF THE CASE STUDY

The analysis of the case study in this section is from the data collected at previous section, i.e., Section 3 of our study. We used three separate Google forms to gather data from Computer Science (CS) students, Information Technology (IT) / Cybersecurity students, and non-CS, non-IT, non-CyberSec students. We have compared the data in the Table-1 below of all three Forms (Form 1 - IT/CybSec, Form 3 - CS, Form 2 - non-CS/non-IT/non-CybSec) and then plotted as the Bar graph of the data after the Table I.

TABLE I. COMPARISON & ANALYSIS OF UNIVERSITY STUDENTS' RESPONSE OF TEN SURVEY QUESTIONS FOR THREE MAJORS

| QN | Questions | CS (Form 3) | non-CS/non-IT/non-CybSec (Form 2) | IT/CybSec (Form 1) |
|---|---|---|---|---|
| 1. | How often do you backup important data and files? [a. Daily b. Weekly c. Monthly d. Never] | Monthly: 40% Daily: 36% Weekly: 24% Never: 8% | Daily: 36% Weekly: 32% Monthly: 24% Never: 8% | Monthly: 52% Never: 20% Weekly: 16% Daily: 12% |
| 2. | Do you use two-factor authentication for imp accounts *(e.g., banking, email)*? [a. Yes b. No] | Yes: 88% No: 12% | Yes: 76% No: 24% | Yes: 96% No: 4% |
| 3. | Have you ever downloaded a file/program from an untrusted source? [a. Yes b. No c. Maybe] | Yes: 56% No: 20% Maybe: 24% | Yes: 24% No: 28% Maybe: 48% | Yes: 48% No: 32% Maybe: 20% |
| 4. | Do you regularly update the software and OS on your devices? [a. Yes b. No] | Yes: 84% No: 16% | Yes: 84% No: 16% | Yes: 88% No: 12% |
| 5. | How many hours (approximately) do you use the internet daily? [a. 1-5 b. 6-10 c. 10+] | 6-10: 60% 1-5: 24% 10+: 16% | 6-10: 48% 1-5: 44% 10+: 8% | 1-5: 44% 6-10: 32% 10+: 24% |
| 6. | Is cybersecurity the responsibility of individuals or organizations? [a. Individual b. Organization c. Both] | Both: 84% Individual: 8% Organization: 8% | Both: 80% Individual: 16% Organization: 4% | Both: 84% Individual: 4% Organization: 12% |
| 7. | How do you pay your utility bills *(phone, gas, electricity, water, etc.)* ? [a. Online b. Offline] | Online: 84% Offline: 16% | Online: 88% Offline: 12% | Online: 80% Offline: 20% |
| 8. | Did you attend online class as part of your degree program? [a. Yes b. No] | Yes: 92% No: 8% | Yes: 52% No: 48% | Yes: 88% No: 12% |
| 9. | Do you use Wi-Fi in public places *(shopping malls, restaurants, airports etc.)*? [a. Yes b. No] | Yes: 80% No: 20% | No: 72% Yes: 28% | Yes: 60% No: 40% |
| 10. | How often do you change your passwords? [a. Weekly b. Monthly c. Yearly d. Rarely] | Rarely: 48% Monthly: 32% Yearly: 20% Weekly: 0% | Monthly: 52% Rarely: 36% Weekly: 8% Yearly: 4% | Rarely: 52% Monthly: 24% Yearly: 24% Weekly: 0% |

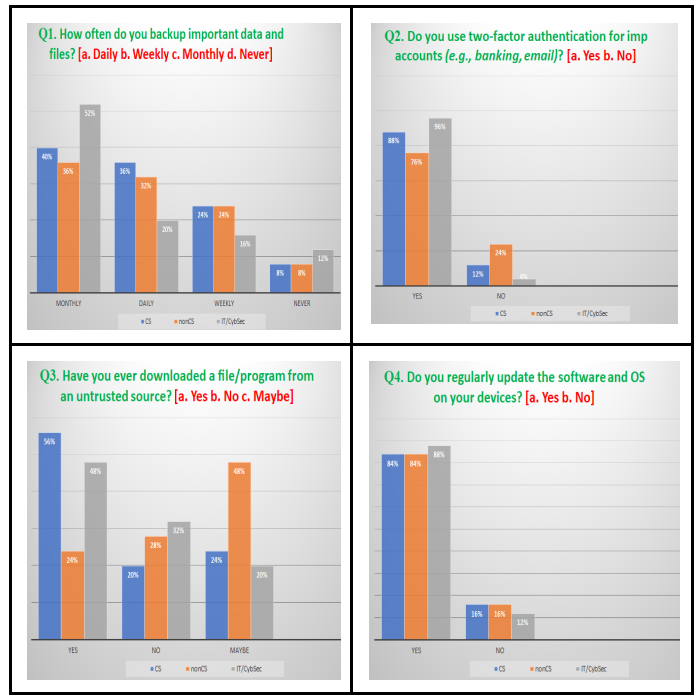

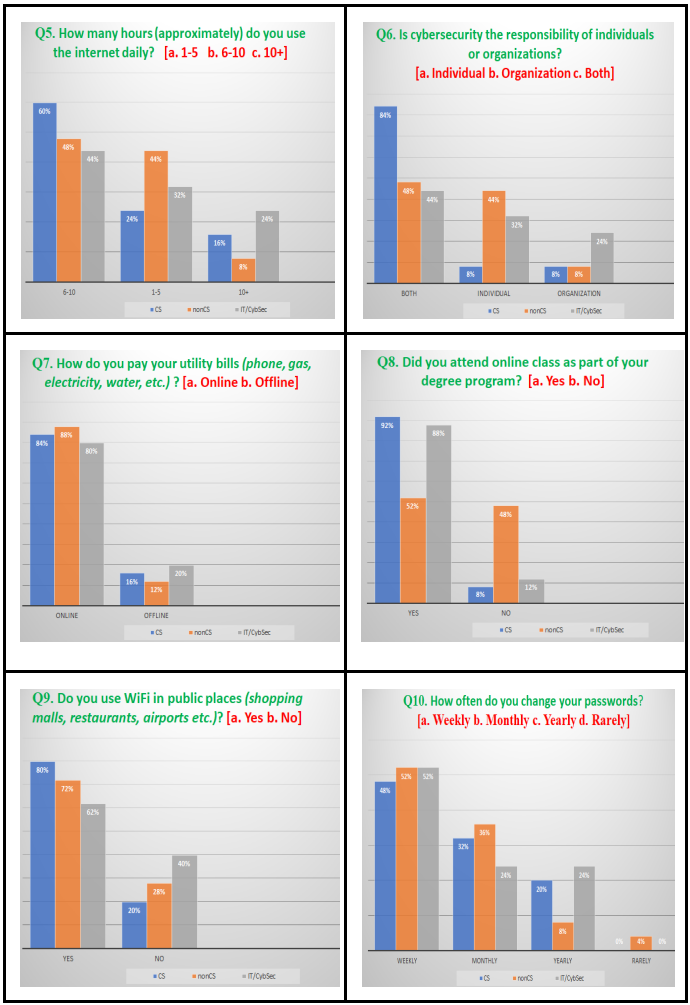

Fig. 1. Bar graph of the University students' data for ten survey questions for three majors

## V. NEED OF CYBER AWARENESS

There are several factors that cause cyber awareness as an immediate need in today's cyber world to develop cybersecurity education and programmes [14]. We have explored and also explained the needs below for cyber awareness:

A. Rise in online and distance education system worldwide [2] [15]: Online and distance learning is an important way to achieve knowledge, degrees, certificates in the post-COVID era. Number of students globally accessing online, and distance education courses as compared to traditional learning exceeds pre-COVID levels with top three countries as United States, India, and Mexico according to Coursera. The mass of students did not have satisfactory knowledge about internet use and cyber threats. Cybersecurity awareness training can not only tutor the students to be equipped for possible cyber threats but also notify them about the legal facet of cyberattacks and cybercrime.

B. Many cyber threats in social media than before [3] [16]: Since January 2021, social media threats targeting enterprises have increased 47% as per the report of PhishLabs. Fraud related to online payment services, and impersonation attacks are the major two social media threats discovered apart from data leakage, cyber threat, and physical threat. Social media is a pool of e-communication platform used by online users to allocate information, ideas, personal messages with each other. Most social media operators and online students are uninformed of the risks and vulnerabilities coupled with social platforms unless experiences in their real lives.

C. Maximum use of data, internet, and instant messaging [5] [17]: Data usage by apps in cellular network when not connected to our unlimited Wi-Fi is maximum now. Voice and video calling apps also consume a lot from data, public use messaging apps to protect money and dodge mobile bills as much as possible. Some free messaging app that consumes 40% - 45% less data compared to others are Pinngle, Skype, Viber for instant messaging. Entrance to information has become a lot simpler than before but novel types of cybersecurity threats typically results in data damage and information abuse have emerged dramatically where students are engaged in data breaches in most cases.

D. Various forms of phishing attacks and fake advertisement [6] [18]: The key to prevention and protect our data from phishing emails is to create a high level of cybersecurity awareness through training and practice that includes careful email reviews, pause, and analyze before automatically clicking "reply", downloading unsecure attachments. Phishing attacks including fake advertisement have evolved over the years that happens when a victim acts on a suspicious email demanding urgent actions. To avoid from a victim of cybercrime all must know about their private security and safety measures to defend by themselves. Protecting the confidentiality of information in complex network systems is very challenging today.

E. No or less knowledge of protection tools and training programs [8] [19]: Cyber awareness and cybersecurity training for small businesses have several benefits including protection of sensitive information, reduced costs, boost employee productivity, maintain customer trust, and increased security awareness. As data usage charges and internet utilization continue to upsurge, cyber awareness is increasingly critical, and several customers including university students have less knowledge and benefits of cybersecurity training programs. Higher cyber knowledge, protection tools is tied to the level of cyber awareness, outside the differences in gender or nation.

F. Vulnerable to cyber-attacks and no cyber defense [9] [20]: All the information and data presented in web are extremely vulnerable to cyberattacks where it lacks transparency and trust. To manage technology roles, minimize corruption, maintain privacy, satisfy citizens including students, new emerging technology such as blockchain can be used. Cybersecurity has arisen as a crucial concept in everyday life as internet-based attacks have become frequent and are projected to increase as technology ubiquity upsurges. Cyber awareness and blockchain technology are now a key defense in the security of people and systems.

G. Environment-dependent awareness and cybersecurity behavior [10] [21]: In several ways, cybersecurity is about behavior of attackers, social engineering methods they employ, tactics used to avoid security detection, trick users within an organization. Security awareness is the major key to cybersecurity behavior change. Humans are the vital figures in cyber security and way to lessen risk in cyberspace is to make people more safety aware. There are various aspects of cybersecurity awareness, they are mutually inconsistent and environment-dependent.

H. Protect corporate data and services via cybersecurity education [13] [22]: Cybersecurity education is an important aspect today, and blockchain mounts as a solution in cyber education while continuing trust and tamper-proof records management. The blockchain platform as Smart contract, Hyperledger, Ethereum, Digital storytelling, Consortium, Distance learning with features of transparency, distributed ledger, privacy, accessibility, interoperability, decentralization, non-intermediaries, traceability, and immutability can protect corporate data and information including services. Existing programmes can also be room for improvement while developing cybersecurity education and awareness programmes.

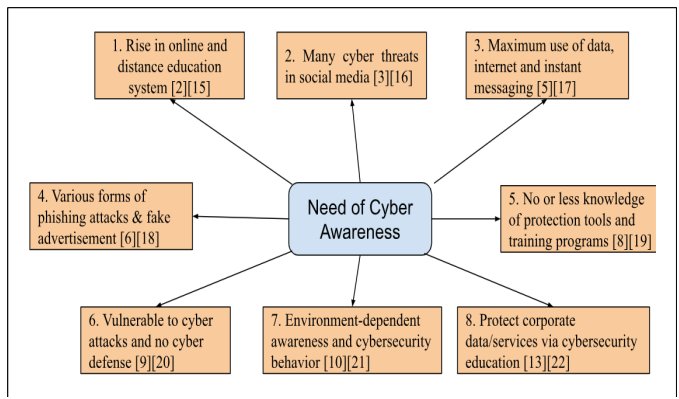

Fig. 2. Need of cyber awareness based on the above Literature Review

## VI. CONCLUSION

We have made several conclusions from the above presented research and case study. The first and foremost thing is that more than 50% of the student back up their data only once a month (CS: 40%, IT/CybSec: 52%. non-CS/non-IT/non-CybSec: 24%), which is serious concern of cyber threats and cyber-attacks because our data are attacked every seconds online. Many student still do not use two-factor authentication (CS: 12%, IT/CybSec: 4%, non-CS/non-IT/non-CybSec: 24%), which is another gateway for cyber-attacks and enforced by several organizations to minimize cyber-attacks. Students also confirmed (with Yes and Maybe options) that they downloaded file/program from an untrusted source (CS: 80%, IT/CybSec: 68%, non-CS/non-IT/non-CybSec: 72%), which is another boost to cybercriminals. Students also do not update the software and OS on their devices (CS: 16%, IT/CybSec: 12%, non-CS/non-IT/non-CybSec: 16%) beside several reminders and warnings. The rise in use of internet daily exceeding 10+ hours after the COVID scenario (CS: 16%, IT/CybSec: 24%, non-CS/non-IT/non-CybSec: 8%) is another ground for cyber-attacks as many students/employees work from home now. Many students still believe that cybersecurity is the responsibility of organizations only (CS: 8%, IT/CybSec: 12%, non-CS/non-IT/non-CybSec: 4%), which is very sad. More than 80% of students now pay their utility bills online (CS: 84%, IT/CybSec: 80%, non-CS/non-IT/non-CybSec: 88%) in the post COVID era, that ultimately increases online transactions and online banking. Also, many educational institutions, colleges, and universities went online so, many students attended online class as part of their degree program than before (CS: 92%, IT/CybSec: 88%, non-CS/non-IT/non-CybSec: 52%). The massive use of public Wi-Fi in public places like shopping malls, restaurants, airports, (CS: 80%, IT/CybSec: 60%, non-CS/non-IT/non-CybSec: 72%) also boils the cyber space and cyber scenario. The last option related with two-factor authentication is changing the passwords at frequent time-intervals, despite enforced by many organizations and institutions, students rarely or yearly changed their passwords (CS: 68%, IT/CybSec: 76%, non-CS/non-IT/non-CybSec: 40%). The non-CS/non-IT/non-CybSec students seems more cyber aware and knowledgeable than CS and IT/CybSec students from the case study presented. The non-CS/non-IT/non-CybSec students (8%) changed their password weekly whereas 0% CS students and 0% IT/CybSec students changes their password weekly. Also, 48% of non-CS/non-IT/non-CybSec students do not attended online class and only 8% non-CS/non-IT/non-CybSec students spend 10+ hours daily on internet with 88% non-CS/non-IT/non-CybSec students paying utility bills online.

## VII. FUTURE SCOPE

There are several weaknesses and limitations of the presented case study and research. Our case study and analysis was limited to university students only (not with academic employees, security professionals, cybersecurity job holders, and others) with 25 each in all three categories making a total of 75 entries. A second limitation of the presented study is that we only used primary data collection from Google Forms (in Questionnaire format only) but did not include a secondary option for data collection procedure. Another third limitation is that we developed exactly ten questions for data collection in response to knowledge and awareness of university students only. Hence, future research is needed to explore more (by overcoming the stated limitations in this study) in terms of cybersecurity knowledge and cybersecurity awareness for an individual. Another fourth limitation might be the bitter truth that the data collection procedure was done in the online format with the help of an online Google Form questionnaire leaving the possibility of obtaining different results if there was a face-to-face survey methodology.

**Authors Contributions** Conceptualization: NK, Methodology: NK, DH, VD, Validation: NK, DH, VD, Investigation: NK, DH, VD, Writing: NK, DH, VD, Supervision: NK, Review & improvement: NK, VD. All authors have read and agreed to the published version of the manuscript.
**Conflict of Interest** The authors declare no conflict of interest regarding the research work.
**Acknowledgement** We would like to acknowledge everyone who helped us directly and indirectly in conducting this research.
**Data Availability Statement** The data presented in this study are available upon request from the corresponding author.
**Funding** This research was partially funded by a grant (Teaching Excellence & Innovation Grant 2022) at Lindenwood University, MO, USA.